\documentclass[preprint]{aastex}

\usepackage{emulateapj5}
\usepackage{apjfonts}

\usepackage{epsfig}

\newcommand{\VI}{$V_{606}-I_{814}$\,}
\newcommand{\JH}{$J_{110}-H_{160}$\,}
\newcommand{\V}{$V_{606}$\,}
\newcommand{\I}{$I_{814}$\,}

\begin{document}

\title{The Relative Star Formation Histories of Spiral Bulges and
Elliptical Galaxies in the Hubble Deep Fields}

\author{Richard S. Ellis\altaffilmark{1}}
\affil{Department of Astronomy, California Institute of
Technology, Pasadena, CA 91125. rse@astro.caltech.edu}

\smallskip

\author{Roberto G. Abraham\altaffilmark{1}}
\affil{Department of Astronomy, University of Toronto, 60 St.
George Street, Toronto, ON M5S3H8, Canada.
abraham@astro.utoronto.ca}
\medskip
\and

\author{Mark E. Dickinson}
\affil{Space Telescope Science Institute, 3700 San Martin Drive,
Baltimore,MD 21218. med@stsci.edu}

\altaffiltext{1}{Also: Institute of Astronomy, University of
Cambridge, Madingley Road, Cambridge, CB3 OHA, United Kingdom}

\begin{abstract}

Hierarchical galaxy formation models make specific predictions
concerning the relative assembly rates and star formation
histories of spiral bulges and field ellipticals. Independently of
the cosmological model and initial power spectrum of fluctuations,
at all epochs the stellar populations in spiral bulges should be
older and redder than those in typical ellipticals selected at the
same redshift. To test this simple prediction, we analyze the
internal optical colors of a complete sample of \I $<$24 mag
early-type and spiral galaxies from the Northern and Southern
Hubble Deep Fields (HDF). The subset of galaxies in the Northern
HDF are also investigated in the near-infrared using NICMOS
photometry. We compare the central (inner 5\% radius) colors of
those spirals with clearly visible bulges with the integrated
colors of ellipticals in our sample. Comparisons are possible to a
redshift $z\simeq$1 at which point well-defined bulges become
difficult to locate. The reliability of determining bulge colors
using central apertures is tested by considering the homogeneity
of the pixel-by-pixel colors for typical cases and through
comparisons based on the simulated appearance at moderate redshift
of the local sample of de~Jong. We show via these tests and by
selecting HDF subsets chosen according to inclination that disk
contamination effects should be minimal. While spiral bulges are
systematically redder in their optical colors than their
associated disks at all redshifts,  we find that the majority are
significantly bluer than the red locus occupied by most field
ellipticals at similar redshifts. In the near-infrared, similar
trends are found at redshifts $z<0.6$, but at higher redshifts
some bulges as red as the reddest ellipticals are found. We
conclude that a significant rejuvenation may have occurred in the
inner stellar populations of many spiral galaxies, particularly
those at intermediate redshifts. We examine the optical and
near-infrared colors of the HDF bulges in the context of models
which include the effects of secondary star formation superimposed
upon pre-existing old populations and conclude the data is best
fit when this secondary activity is burst-like. We discuss the
consequences for models of secular evolution in disks should these
bursts have been particularly prevalent at $z\simeq$0.6 as the
limited HDF data seems to imply.

\end{abstract}

\section{Introduction}

The origin of galactic bulges is an important and unsolved issue
in our understanding of galaxy evolution. The hitherto
established view is that galactic bulges form at high redshifts
through early dissipationless collapse (Eggen, Lynden-Bell, \&
Sandage 1962, hereafter ELS). This is based principally on the
evidence for old stellar populations concentrated in the bulge of
our own Galaxy and contrasts with more recently-developed
hierarchical galaxy formation models (Kauffmann et al 1993, Baugh
et al 1998) where elliptical galaxies form from the merger of
early disk systems which can, in turn, continue to accrete gas to
form a two component spiral galaxy.

A robust prediction of all hierarchical models is that spiral
bulges should, on average, contain older stars than their
associated disks which form by subsequent accretion. Moreover,
statistically at a given redshift, bulges should be older and
redder than field ellipticals which predominantly form from the
merger of previously created spirals. Importantly, these
conclusions should remain valid regardless of the particular
cosmological model or initial power spectrum which governs the
rate of assembly of massive galaxies. As such, a comparison of the
relative colors of ellipticals and spiral bulges offers a
remarkably simple, but powerful, test of hierarchical assembly
models.

A third alternative for the origin of stellar bulges proposes
their manufacture via various instabilities of pre-existing disks
(see the recent review by Combes 1999). Of greatest interest here
is the suggestion that bulges may form through secular processes
(those whose timescale are longer than the dynamical time), mostly
through interactions and the evolution of galactic bars. Galactic
bars can be transformed into bulges through a vertical heating of
the inner disk via resonant scattering of stellar orbits by the
bar potential. Numerical work also suggests that bulges may form
via bar destruction through the growth and subsequent collapse of
bar instabilities in cold rotating disks. Evidence that such
secular activity may play an important role in the formation of
bars is growing (Kormendy 1993; Kuijken \& Merrifield 1995;
Norman, Sellwood, \& Hasan 1996) and there is rather good
dynamical evidence that most bulges in very late-type galaxies
are not true bulges, but rather vertically heated central disks
(Kormendy 1993). This suggests that the relative importance of
the three bulge formation processes may be a strong function of
Hubble type.

Most investigators of bulge formation have sought to infer their
origin by studying the stellar populations or dynamical properties
of local galaxies. Of particular importance in this regard are
extinction-free near-infrared imaging observations of the Milky
Way bulge (Dwek et al 1995) and resolved studies of associated
stellar populations (eg. Rich 1997). These studies and those
possible for nearby galaxies have emphasized a much greater
diversity in color and stellar history than had hitherto been
believed (Wyse et al 1997) although the effects of dust and other
factors remain unresolved (Peletier et al 1999).

Through the Hubble Deep Field (HDF) campaigns (Williams et al
1996, 1999), high resolution imaging data is now available with
sufficient signal-to-noise to enable the direct study of bulge
growth and evolution {\em in situ} to redshifts $z \sim 1$. Given
the above discussion, it is clear that a detailed observational
comparison of the resolved stellar populations of field spirals
and ellipticals as a function of redshift will offer important
insight into some of the most fundamental aspects of galaxy
formation. Indeed, only via intermediate redshift data can the
epoch corresponding to the formation of the bulge component be
determined independently of the age of its stars.

In an earlier paper (Abraham et al. 1999a) we investigated the
importance of secular processes in forming bulges by determining
the proportion of barred spiral galaxies in the Hubble Deep Fields
as a function of redshift, finding a tentative decline in the
fraction of barred galaxies beyond redshift $z\simeq0.5$. The
physical mechanisms responsible for such a deficit of barred
spirals at high redshift remain unclear. Possibilities include
dynamically hotter (or increasingly dark-matter dominated) disks
or an enhanced efficiency in bar destruction at high redshift.
Regardless, the absence of bars at high redshift does seem to pose
an observational challenge for models where bulges form via
secular activity initiated by bars. Such activities may be
episodic (Sellwood 1999), or confined to low-luminosity late-type
spirals (Kormendy 1993) which are too faint at high redshifts to
have been included in our earlier investigation.

In a precursor of this study (Abraham et al. 1999b), the
dispersion in the internal colors of a sample of high-redshift HDF
spirals with redshifts was used to investigate the relative colors
and star-formation histories of bulges and disks at high
redshifts. This work concluded that bulges are redder and older
than their surrounding disks in essentially all luminous
high-redshift spirals. However, while bulges appear redder than
their surrounding disks out to high redshifts, no investigation of
the relative colors of bulges and ellipticals as a function of
redshift was undertaken at that time.

Such a comparison forms the basis of this paper. Here we continue
our HDF investigations concerning the origin of galactic bulges by
measuring the relative colors of spiral bulges and ellipticals as
a function of photometric and spectroscopic redshift. Using
automated morphological classifications, our study includes a
total sample of 95 spirals and 60 early-type galaxies with \I
$<$24 mag in the Northern and Southern Hubble Deep Fields. An
additional aspect of this work is the augmentation of our optical
data by near-infrared NICMOS observations (Dickinson 2000) of a
subset of our data, extending coverage into the $J_{F110W}$ and
$H_{F160W}$ photometric bands in the Northern Hubble Deep Field.
These infrared observations are neither as deep nor as
well-sampled as the optical data in the HDF-N, but provide a
number of interesting constraints on the conclusions derived from
optical photometry alone.

A plan of this paper follows. In \S2 we describe our overall
sample selection and the techniques used to assign redshifts to
these galaxies. In \S3 we describe our technique for estimating
the relative colors of ellipticals and bulges via the measurement
of a central aperture \VI color. We quantify the uncertainty in
these estimates for the bulges in our sample by examining
individual cases and applying a similar technique to the local
spiral galaxy sample of de~Jong (1996). We discuss the
bulge-elliptical comparison in the context of hierarchical models
in \S4 and summarize our main conclusions in \S5.

\section{The HDF Sample}

As in Abraham et al (1999a,b), we take advantage of the unique
nature of both Hubble Deep Fields (HDFs, Williams et al 1996,
1999). The long exposures in these observations guarantees
substantially improved signal to noise in several passbands over
earlier multi-color HST imaging surveys such as the Medium Deep
Survey (Ratnatunga et al 1999). This is particularly important in
the reliable separation of ellipticals from spiral bulges with low
surface brightness disks. The 0.04 arcsec/pixel sampling of the
drizzled optical data makes isolation of the bulge comparatively
easy, as described below (\S3). By combining both HDFs, a larger
sample is produced which is essential for statistical comparisons
such as the color evolution of bulges and ellipticals at a given
redshift.

For the Northern HDF field, we supplement our optical data with
near-infrared NICMOS data. (The characteristics of these data are
described in Dickinson 2000, and are only summarized here.)
Although the HDF-N was observed in the near-infrared from the
ground in several programs, the depth and angular resolution of
these data ($\simeq$1 arcsec FWHM) are a poor match to that of
the optical WFPC2 images. The present NICMOS HST data was
obtained by mosaicing the complete HDF-N with a mean exposure of
12600s per filter in F110W (1.1$\mu$m) and F160W (1.6$\mu$m). The
sensitivity varies over the field of view but the mean depth is
$AB\simeq$26.1 at 10$\sigma$ within a 0.7 arcsec diameter
aperture. The drizzled NIC3 PSF has a FWHM = 0.22 arcsec and is
primarily limited by the NIC3 pixel scale.

The \I =24 mag limit adopted for the present morphological study
is a conservative one. For example, Ratnatunga et al (1999) claim
reliable morphological classifications are possible to \I =26 mag
in the northern HDF. An important consideration here is the
reliable isolation of the bulge component at high redshifts. We
defer discussion of this point to \S3. However, we note at this
point that bulges becomes difficult to reliably detect beyond a
redshift $z\simeq$1. As a significant fraction of galaxies beyond
\I $>$24 have redshifts beyond unity, this is an additional
justification for our adopted magnitude limit.

The baseline photometric catalogs for the present study are the
publicly available {\tt SExtractor} optical catalogs produced by
the Space Telescope Science Institute (STScI). Throughout the
present paper, we will adopt IAU-format object designations and
coordinates keyed to these catalogs. To $I_{814}=24$ mag, both
HDF images contain 242 galaxies of all types. Morphological
classification of this sample was undertaken using the automated
classification technique (based on measures of central
concentration and rotational asymmetry) described in Abraham et
al. (1996a,b), supplemented by visual classifications determined
by two of the authors (RSE and RGA). This is the same strategy
adopted in our analysis of bar structure in the HDF fields
(Abraham et al 1999a). The galaxies in our sample are
sufficiently bright that there is no ambiguity in matching
galaxies to their counterparts in the corresponding NICMOS image
of the northern HDF.

Table 1 summarizes the properties of the full sample. The table
records integrated and central optical and near-IR colors as
defined later in the paper, quantitative structural measures
(central concentration, rotational asymmetry, axial ratio, and
automated classification), and visual classifications according to
two of the authors (RSE and RGA). Visual classifications are on
the MDS numerical system: -2=star, -1=compact, 0=E, 1=E/S0, 2=S0,
3=Sab, 4=S, 5=Scdm, 6=Ir, 7=peculiar, 8=merger, 9=defect. A
measure of the uncertainties for the purpose in hand (selecting
spirals and ellipticals) can be gained from comparisons of both
visual classification schemes with that determined automatically.

Automated classification results in a sample of 95 spirals and 60
early-type systems, the remainder being irregulars, mergers or
peculiars. Visual classification by RSE (RGA) results in a total
of 96 (123) spirals and 74 (64) early-types. In the present paper
visually classified intermediate type systems, ie. S0/a galaxies,
were categorized as early-type. (Nine galaxies in total were
classed as S0/a systems, and we will demonstrate in \S4 that our
results do not depend sensitively on how they are partitioned).

The RSE classes are in excellent agreement with those determined
automatically for spirals but a 23\% excess of ellipticals is
found. Conversely, RGA finds good agreement with the automated
scheme for ellipticals but has a 29\% excess of spirals. This
level of discrepancy is consistent with earlier comparisons
between independent visual classifications of faint galaxies
(Abraham et al. 1996b). Concerning the apparent discrepancies for
spirals, it should be noted that the automated classifier is tuned
to categorize weakly distorted systems and very late-type spirals
as peculiar. The excess of visually-classified ellipticals is
thought to arise because some visual classifiers weight global
symmetry more than central concentration (Marleau \& Simard 1999).

The most important consideration here is the selection of reliable
samples of ellipticals and well-defined spirals (the latter from
which the bulge properties can be determined). In the case of the
ellipticals, we follow precisely the selection criteria of Abraham
et al (1996,1999a) for the HDF-N and its equivalent procedure for
the HDF-S (Menanteau et al 2000). Discussion of possible
contamination of these elliptical samples by compact galaxies can
be found in Menanteau et al (2000). In the present paper no
attempt is made to separate S0s from ellipticals.

As described below, we will restrict our spiral sample to those
non-peculiar systems with prominent bulges. For this subset both
visual and automated classifications give virtually identical
results so far as locating spirals is concerned. Regular spirals
and those with visible bulges were ascertained by visual
inspection by both observers. However, one of us (RGA) examined
those spirals he classed in more detail for blending effects and
these more detailed visual morphological classifications
(summarized in Table~1) form the basis of the remainder of the
analysis. The principal goal here is to enable selection of
various (presumed ``clean'') subsets of spiral bulges for later
analysis. We will show in \S4 that the conclusions of this paper
do not depend sensitively on the exclusive use of any one of these
subsets.

For the RGA-classed spirals, the column denoted ``{\tt bulge?}''
indicates the presence of a bulge component. The column denoted
``{\tt blend?}'' indicates whether nearby companions or optical
superpositions overlap with a galaxy image at a 2$\sigma$
isophotal threshold on the \I-band image. Blends are categorized
as either ``severe'' or ``minor''. Severe blends are cases where
the overlapping images cannot be reliably disentangled. The
distinction between a ``severe'' blend and a merger is obviously
rather subjective. In some cases tidal tails make the latter
classification obvious, while conversely ``severe blending'' is
appropriate when contamination originates from, {\em e.g.} a
diffraction spike presenting photometric difficulties. The
``minor'' blend category encompasses systems where the overlapping
images can be cleanly separated by raising the $2\sigma$ isophotal
threshold, used to define the galaxy above the sky background, by
a small amount (typically an additional 1--3$\sigma$ above the sky
noise level). The tabulated photometric parameters for galaxies
flagged as minor blends in Table~1 correspond to the properties
determined at this higher threshold. Throughout the remainder of
this paper these minor blends will be treated in the same manner
as isolated galaxies. The flag, ``{\tt pec?}'' indicates the
visual presence of an optical peculiarity that is insufficient to
warrant classifying the galaxy as a peculiar, {\em eg.} a spiral
that could reasonably be classed as Spec in local catalogs.

Figure~1 (Plate 1) presents a montage, ordered by \I magnitude,
of the 68 isolated non-peculiar systems classified as spirals by
RGA whose bulge components are clearly visible, ie. systems
selected from Table~1 using the following criteria: {\tt (RGA =
Sab, Sbc, or Disk)},\  {\tt bulge?~=~yes},\ {\tt pec?~=~no}, and
{\tt blend?~$\ne$~severe}. We note that the prominent bulges and
fairly tight spiral structure visible on many of the spirals
shown Figure~1 indicates that the majority of systems are earlier
than type Scd. This is of course consistent with our initial
pre-selection designed to isolate spirals with prominent bulges.

Galaxy redshifts are listed on each panel of the montage.
Spectroscopic redshifts become progressively fewer fainter than
\I=22.5 mag in the HDF-N and are still rather scarce in the HDF-S.
We have therefore augmented the spectroscopic redshift estimates
with photometric redshifts based on the publicly available catalog
produced by Gwyn ({\tt \small
http://astrowww.phys.uvic.ca/grads/gwyn/pz/index.html}). These
photometric redshifts are based on optical photometry only and do
not include broad-band infrared colors. However, the summary of
results in Hogg~et~al. (1999) suggests that infrared photometry
does not significantly improve the accuracy of the photometric
redshifts in the HDF for the $z<1$ range of importance in this
work.

\section{Photometric Decomposition}

Many studies of high-redshift bulges have attempted to decompose
galaxian light into components by fitting analytical models to the
surface brightness distributions of galaxy images (Schade et al
1995, Lilly et al 1998, Marleau \& Simard 1998, Ratnatunga et al
1999). In this procedure, all pixels in the image are used to
determine simultaneously the global structural parameters for all
components of the galaxy, such as the scale lengths,
ellipticities, and characteristic surface brightnesses for the
bulge and disk. By integrating over the results from such fits,
the colors of galactic components can be inferred.

Obtaining robust model fits using this technique can be
challenging and, at a more fundamental level, the approach assumes
the validity of canonical fitting laws and a close correspondence
between stellar populations and global galactic structure (a
relationship that is likely to vary reasonably strongly with
Hubble type). Both of these may be poor assumptions, particularly
for later Hubble types (Kormendy 1993). By utilizing all pixels in
the image to determine model parameters this technique clearly
optimizes signal-to-noise, but of course this advantage can also
prove a liability, as a small number of pixels affected by
features not incorporated into the models (such as dust) can have
a global impact on the fits. Furthermore, the morphological
irregularity of high redshift galaxies suggests that fitting
smooth light distributions to distant galaxies may introduce
spurious results if care is not taken to first isolate underlying
smooth features from superposed asymmetrical structures. This idea
is very difficult to implement in a scale-free manner, although
some attempts based on enforcing rotational symmetry seem quite
encouraging (Schade et al. 1995; Lilly et al. 1998), and
alternative fitting methods which attempt to minimize the effects
of irregular features by adopting particularly robust minimization
strategies (eg. simulated annealing) also show promise (Marleau \&
Simard 1998).

In the present paper, we choose to adopt a simpler approach which
foregoes any attempt to determine the global structural parameters
of the galaxies in our sample, in order to focus specifically on
bulge stellar populations in a robust non-parametric manner. We
estimate bulge colors by measuring the colors of the innermost
regions of galaxy images where the bulge component dominates. This
approach is conceptually similar to that adopted by Peletier et
al.  (1999) in determining bulge colors for local spirals using
HST NICMOS observations, although the details (described below)
differ.

Ideally an aperture selected to locate the bulge should be defined
so as to be linked physically to the intrinsic properties of the
galaxy. In this way correlations with, e.g. redshift, disk surface
brightness and luminosity would be minimized. However, such an
approach suffers from many of the difficulties associated with a
full decomposition (see above) and, in practice, when observing
with an instrument (WFPC-2) with a fixed pixel size, turns out to
be very difficult to arrange. A metric aperture cannot easily be
chosen to simultaneously avoid disk contamination at low z whilst
containing an adequate number of HST pixels (from signal/noise
considerations) at large redshift.

Instead we adopted an aperture size which scales with the
isophotal size of the galaxy. This has the merit of great
simplicity, the argument being that any systematic effects can be
explored through careful simulations based on nearby galaxies. A
potential drawback of this approach is that the nuclear colors so
defined may be contaminated by disk light in a manner that is a
function of redshift (assuming disks are generally bluer than
bulges at high redshifts, cf. Abraham et al. 1999b). We show below
from detailed simulations that, for the nuclear regions of
low-inclination spirals with morphological $T$-types $T<6$ (ie. Sc
and earlier), bulge light dominates over disk light out to quite
high redshifts in both \V and \I filters, so that the {\em
central} \VI color traces bulge colors with an RMS accuracy of
$\sigma_{V-I} < 0.15$ mag out to $z<0.7$.

Further tests we introduce to examine possible systematic effects
are the analysis of subsets of our sample of spirals with
prominent bulges defined in the previous section, restricted
according to disk inclination and comparisons between optical and
optical-infrared colors since disk contamination should be minimal
in the latter case.

\subsection{Methodology}

In this section we describe the procedure used to analyze our
primary WF/PC2 optical dataset.  The analysis of our supplementary
NICMOS observations of the HDF-N field proceeded along essentially
identical lines.  However, it is important to make clear at this
stage that the under-sampling of the NIC3 detector relative to
WF/PC2 observations, and the rather uncertain photometric
calibration of the NICMOS instrument, make it substantially
harder to apply the techniques described in this section to the
NICMOS observations. For these reasons (and because the NIC3 data
are restricted to HDF-N), in the present paper we will treat the
NICMOS data as an interesting adjunct to our optical data that is
mostly useful for testing and elaborating the conclusions derived
at from data obtained at optical wavelengths, and defer further
discussion of the NICMOS data until \S5.4.

Our analysis is based on elliptical apertures defined using second
order moments obtained from the $I_{814}$-band image of each
galaxy in Table~1. These were used to define an ellipse whose area
corresponds to that of the galaxy above a 2$\sigma$ isophotal
threshold. Large aperture (`total') colors (see Table~1) were
defined using this procedure applied to the $I_{814}$ image. For
each galaxy a circular aperture was then defined with a radius of
5\% of the semi-major axis length of the galaxy's ellipse, and
this inner aperture was then used to determine the \VI colors of
the central portion of the galaxy.

The 5\% isophotal radius aperture size was chosen according to the
following considerations. Clearly the aperture should be as small
as possible in order to isolate bulge light with a minimum of
contamination from disk light. However the aperture must be large
enough to encompass several pixels from signal/noise
considerations. For a minimum of 10 pixels to be averaged, the
aperture radius must be at least 2 pixels corresponding to an
angular size of $\simeq$0.08\arcsec using the dithered 0.04\arcsec
pixels of the Version~2 HDF images released by STScI. Assuming
$H_0$=70 km sec$^{-1}$ Mpc, $\Omega_M$=0.3 and
$\Omega_\Lambda$=0.7 (the cosmology adopted throughout the paper),
0.08\arcsec corresponds to 0.64 kpc at $z$=1.

The location of each aperture was automatically adjusted by the
measurement algorithm so as to be positioned on the local centroid
of the galaxy image. The central and total \VI colors for all
galaxies in our sample (including ellipticals) are given in
Table~1. The photometric zero points are given in Williams et al.
(1996).

Initially we considered defining the bulge interactively for each
galaxy by ``growing'' the bulge region out from the nucleus on a
pixel-by-pixel basis, isolating the inner pixels of the galaxy
whose resolved pixel-by-pixel color distribution was homogeneous
at some pre-determined level. This methodology was found to work
well for bright galaxies. However, as was found in Abraham et al.
(1999b), the limiting magnitude for this approach in the HDF data
is around \I=23.2 mag, motivating us to attempt to improve the
signal-to-noise of the measurement by binning together central
pixels using central apertures in order to reach a practical
magnitude limit of \I=24 mag. Thus in the present paper the pure
pixel-by-pixel approach is set aside in favor of a central
aperture. It is important, nonetheless, to understand whether a
5\% aperture radius allows us to chromatically isolate a typical
bulge.

We can test this by examining the pixel-by-pixel optical colors in
a number of bright galaxies as an independent check. Figure 2
shows how the bulge pixel colors are distributed with respect to
those of the outer disk in 3 bright spirals with fractional
aperture radii corresponding to 2\%, 5\% and 7\% of the semi-major
axis length. As can be seen, in all cases the bulge defined by a
5\% radius is remarkably homogeneous and red compared to the
pixels outside. In fact the relative colors of the spiral bulge
and the elliptical in the third and fourth rows (seen at virtually
identical spectroscopically-determined redshifts of $z=1.013$ and
$z=1.016$) foreshadows our major result: the bulge in the spiral
system is significantly bluer than the typical elliptical.

\subsection{Possible Contamination of Bulge Colors by Disk Light}

The suitability of deriving bulge colors using a constant
fractional aperture can be further assessed by considering the
spirals in the local sample of de Jong (1996; hereafter DJ96). In
this section we utilize analytic profile fits to the DJ96 sample
photometry and estimate, on their basis, the likely disk
contamination expected when the DJ96 galaxies are artificially
placed at various redshifts and analysed with our central aperture
technique.

DJ96 presents fundamental disk and bulge parameters in $BVRIHK$
bands for 86 nearby low---intermediate inclination spiral galaxies
as a function of Hubble T-type.  The suitability of any
currently-available local reference sample for the purposes of
comparing with deep HST data is debatable, but the resolution and
rest-frame depth of the data presented in DJ96 corresponds
reasonably closely to that for deep images obtained with HST at
high redshifts. The reader is referred to Lilly et al. 1998 (who
also adopt DJ96 as the local calibration sample for a study of
high redshift spirals) for a further discussion of this point.

It is also questionable whether an $r^{1/4}$-law profile or an
exponential profile provides the most suitable description of
local bulges. Following Frankston \& Schild (1976) and Andredakis
\& Sanders (1994), DJ96 convincingly argue that local bulges are
best fit by $r^{1/4}$ profiles in early-type spirals, and by
exponential profiles in late-type systems, and presents fits based
on both canonical profiles as a comparison. In order to undertake
a more rigorous test of our methodology, and in order to estimate
our measurement errors conservatively, in the present paper we
will adopt the exponential-law profile fits given in DJ96.  The
sharp nuclear spike of the $r^{1/4}$ profile makes determining
bulge colors on the basis of the technique described in the
present paper more accurate.

Figure~3 illustrates the difference between the $U-V$ and $B-R$
bulge colors determined by 5\%-isophotal radius aperture
photometry and those determined by integrating over the DJ96 model
fits. As de~Jong (1996) does not tabulate model fits in $U$-band,
we have assumed a constant rest-frame $U-B=-0.12$~mag for the disk
component, and constant rest-frame $U-B=0.50$~mag for the bulge
component, corresponding to the mean colors for late and
early-type galaxies in Table~1 of van den Bergh~1998. Note that
$U$ and $B$ bands correspond roughly to rest-frame \V at redshifts
of $z\sim 0.4$ and $z\sim 0.7$, respectively, so the left and
right panels of the figure illustrate the expected errors
introduced into our \VI bulge color measurements at these
redshifts. On the basis of this figure, at $z \sim 0.4$ we can
expect the RMS uncertainties on our bulge color estimates to be
$\sigma_{V-I} < 0.1$ for spirals earlier than Sc, and
$\sigma_{V-I} < 0.15$ for essentially all spirals earlier than
type Sdm. At higher redshifts, blue rest-frame disk light may
contaminate the bulge light to a larger degree, but even at $z
\sim 0.7$ the RMS contamination is only $\sigma_{V-I} \sim 0.2$
for spirals earlier than Sc, which we will show below is
substantially smaller than the typical color difference between
ellipticals and bulges at the redshifts of interest. Using NICMOS
\JH colors, disk contamination is expected to be negligible at
$z<1$.

\subsection{Aperture Effects for the Elliptical Galaxies}

In order to facilitate the simplest comparison between the colors
of spiral bulges and ellipticals, we will utilize integrated (i.e.
large aperture) colors for the field ellipticals. As described
earlier, large aperture colors were derived within ellipses
defined by the second order moment of each galaxy in the \I band.
For completeness, Table 1 also lists 5\% aperture photometry for
these sources as this contains important information on the extent
to which ellipticals represent a homogeneous population.

Abraham et al (1999b) found that as many as a third of the HDF
ellipticals to \I = 23.2 have cores bluer than those of the
galaxies' periphery suggesting bursts of star formation involving
at least a few percent of the galactic mass in the previous few
Gyr. Because of this, the use of 5\% aperture colors for the
ellipticals would complicate any comparison with our sample of
spiral bulges.

A complete discussion of the internal optical colors of the HDF
elliptical population is presented in a separate paper (Menanteau
et al 2000). Nonetheless it is interesting to examine the color
inhomogeneities from Table 1 briefly in order to set a scale for
the color variations we later discuss for spiral bulges. For the
HDF ellipticals in the present study, the mean offset in color in
the sense (aperture)-(total) is $\simeq$0.08 mag with $\simeq$15\%
showing no difference at all. Only 10\% of the sample has offsets
greater than 0.2 mag.

\section{Relative Optical Colors of Bulges and Ellipticals}

In this section we will investigate the optical colours of our
complete sample of bulges and ellipticals, and defer consideration
of the infrared subset of these data until \S5.3.

The \VI~colors of ellipticals and spiral bulges as a function of
photometric redshift are compared in Figure~4. This figure also
shows the predicted colors of a simple passively-evolving stellar
population based on the spectral synthesis models of Charlot \&
Bruzual (1999). This baseline model prediction corresponds to a
single 1 Gyr burst at high redshift ($z_F$=3) and assumes solar
metallicity with a Scalo Initial Mass Function. In Figure~4 the
symbol sizes are keyed to central concentration and it is seen
that most spirals in the sample are indeed prominently nucleated
early-types, as expected from our initial pre-selection of
spirals with prominent bulges.

The prominent red locus defined by some (but not all) of the
ellipticals in Figure~4 shows a redshift-dependent trend
consistent with earlier work (Schade et al 1996, 1999, Brinchmann
et al 1998) and, with some exceptions, is in excellent agreement
with our simple passively-evolving model. As discussed by Abraham
et al (1999b), a significant fraction of ellipticals lie blueward
of the well-defined red-locus. Many of these early-type systems
have bluer nuclei and are thought to be remants of recent merging
providing good evidence for continued formation of spheroid
galaxies through mergers of late-type galaxies (Menanteau et al
2000, Brinchmann \& Ellis 2000).

Contrary to expectations, the bulges in our sample of regular,
isolated spirals show a systematic blueward scatter relative to
the red early-type locus. Whereas both bulges and ellipticals show
a dispersion in color at a given redshift, the distributions are
characteristically different. A much smaller fraction of the
elliptical population is blue although there is strong hint of a
bimodality in the distribution, consistent perhaps with bursts or
star formation associated with recent merging (Menanteau et al
2000). By contrast, few of the bulges occupy the `red envelope'
region of the color - redshift relation as expected if they are
intrinsically older than their elliptical counterparts. Moreover
their distribution of colors is less bimodal and occupies the
intermediate region, consistent on the one hand perhaps with less
extreme evolutionary trends but on the other hand ones that are
spread throughout the population.

Before accepting the surprising result that bulges are, as a
population, systematically bluer than ellipticals in the mean, it
is important to investigate possible sources of systematic error.
The main source of systematic error is likely to be contamination
from underlying disk light. As described in \S3.2, the offset
between typical bulge colors and the colors of red ellipticals is
most significant for $z\lesssim 0.5$.  At higher redshifts the
bulk of the effect is unlikely to be explained fully by disk
contamination unless the DJ96 is very unrepresentative of the
high redshift spiral population. Other systematic effects can be
best studied by introducing a measure of the color scatter of the
bulges with respect to our baseline passively evolving
high-redshift collapse model. We define a simple parameter
characterizing the offset between a galaxy's observed color and
the predicted color of our baseline high-redshift burst model:

$$\delta(V_{606}-I_{814}) = (V_{606}-I_{814})_{bulge} -
(V_{606}-I_{814})_{passive}$$

\noindent In Table 2 we present
$\langle\delta(V_{606}-I_{814})\rangle$, the {\em median} value of
$\delta(V_{606}-I_{814})$, and the corresponding RMS dispersion,
$\sigma_{\delta(V_{606}-I_{814})}$, for various subsamples of
spiral bulge populations.

The distribution for the main sample (i.e. that plotted in Figure
4) is compared with that for the ellipticals in Figure 5 clearly
illustrating the points made earlier. For this sample we obtain
$\langle\delta(V_{606}-I_{814})\rangle$=-0.525 and
$\sigma_{\delta(V_{606}-I_{814})}$=0.266. The median colors are
substantially bluer than their elliptical counterparts, and it is
apparent they fill in the regime between the passive elliptical
population and the bluest elliptical outliers. Particularly
striking is the almost complete absence of systems as red as the
passive-evolution track which provides such a good description for
the majority of the early-type population.

Two important conclusions can be drawn from Figure~4 and Table~2.
Firstly, as described in the \S3.2,  we estimate typical
contamination of bulge colors from disk light to be
\VI$\lesssim0.15$ mag for most early---intermediate spirals at
redshifts $z<0.5$. The median offset for the main sample given in
Table~2 is therefore substantially larger ($\simeq\times$3) than
the expected RMS disk contamination for the galaxies in our spiral
sample in this redshift range. At higher redshifts the
significance of the blue bulge colors is hard to determine from
optical data, but Figure~3 suggests that around a third of the
spirals earlier than type Sbc at $z<0.7$ should show little disk
contamination (\VI offsets less than 0.1 mag), although the
optical colors of some systems will be significantly contaminated.

Secondly, in Table~2, the $\langle\delta(V_{606}-I_{814})\rangle$
color offsets and dispersions remain similar to the fiducial
values for our main sample even if broadly different selection
criteria are used to define the galaxy subsamples. Similar results
are obtained when we: (a) restrict consideration only to large
systems where the methodology is expected to be most effective;
(b) restrict consideration only to low-inclination (${b\over
a}>0.5$) systems; (c) restrict consideration only to subsets of
galaxies with confirmed spectroscopic redshifts. {\em The
conclusions that most bulges are systematically bluer than the
reddest ellipticals with a wide dispersion of colors is true
regardless of the detailed prescription used to define the spiral
subsamples.}

\section{Interpretation}

Our principal result is that bulges are, statistically, optically
bluer than the reddest ellipticals and show a large dispersion in
their rest-frame colors. This result is insensitive to disk
contamination, certainly for $z<0.5$ and probably to higher
redshifts since only a very small proportion ($<$5\%) of
intermediate redshift bulges have colors as red as those
ellipticals with occupy a typical passively-evolving track defined
according to a single burst of star formation at high redshift.

In considering the interpretation of these results in this
section, we must take great care to distinguish between three
reasonable working definitions for the age of a bulge component:
(i) optical {\em luminosity-weighted age}, (ii) {\em structural
age} (corresponding to the epoch at which the bulge component is
formed morphologically), and finally (iii) the {\em age of the
oldest stellar population} in the bulge. In the context of our
observations these are not merely semantic distinctions --- in
fact understanding which of these ages is being probed by various
aspects of our data is central to linking our observations with
the predictions of theoretical models.

Our results clearly indicate that the optical luminosity-weighted
ages of bulges to {\em at least} $z=0.5$ are younger than those of
the reddest ellipticals. Restriction of our sample to more robust
(e.g. face-on) sub-samples does not change the strength of this
conclusion. The pixel-by-pixel color distributions of
representative spirals presented in Figure~2 gives a clear
illustration of the overall result.

\subsection{Constraints on Secular Models}

We now consider the implications of younger luminosity-weighted
ages for bulges from the viewpoint of models where bulges grow
through secular evolution (Combes 1999). The paucity of bulges
seen with optical colors close to the evolutionary track for a
single burst of star formation  seriously constrains a dominant
population of old passively-evolving bulges. Similarly, the
optical data is hard to reconcile with the notion that bulge
formation proceeds secularly within the disk {\em without}
associated star-formation in early--intermediate-type spirals.
Further constraints on the secular picture depend upon whether
the star-formation we infer from blue colors is associated with
the formation epoch of the bulge itself, or with ``pollution'' by
young stars upon a pre-existing, older, and morphologically
established bulge component.

In the extreme case, if we assume that the blue colors are
associated with the initial formation of the bulge, then
conceivably {\it some} bulges formed at low redshift from the
secular evolution of stellar disks with bars as an intermediate
stage. Ellipticals could still form continuously from the merger
of young disks at high redshift in which case their predominantly
red colors in the redshift range 0$<z<$1 would be compatible with
growth in a low density or $\Lambda$-dominated cosmology
(Kauffmann \& Charlot 1998, Menanteau et al 1999, 2000). Given the
small proportion of bulges as red as ellipticals in our redshift
range, most bulges would have to have formed surprisingly recently
and one would then expect a significant population of barred
spiral precursors at higher redshift in contrast to the
observations (Abraham et al 1999a). Interestingly, there is some
evidence (c.f. Figure 2) that when bars are present, they do have
redder colors than those of their bulges they contain as expected
in the secular picture. However, numerically it seems unlikely
that most present day bulges were formed in this way unless the
number of $z>$0.5 barred systems has been seriously underestimated
(c.f. Bunker 1999), or bulge formation is episodic, as suggested
by Sellwood~(1999).

\subsection{Constraints on Hierarchical Models}

Next we proceed to discuss the implications of our results in the
context of hierarchical galaxy formation models. Figure~6
illustrates the difference in observed \VI colors expected for
ellipticals, spiral bulges, and S0 galaxies in the $\Lambda$CDM
semi-analytical prescription of Baugh et al. (in preparation), as
a function of both redshift and apparent magnitude. Contrary to
what is observed, the hierarchical picture predicts a much tighter
color distribution for both bulges and ellipticals with the former
being $\simeq$0.1-0.2 mag redder than ellipticals at high
redshifts, and roughly similar colors at low redshift $z\lesssim
0.4$. In contrast to these predictions, we find that at $z<0.5$
(ie. where possible systematic effects introduced by disk
contamination can be neglected) bulges are significantly bluer
than ellipticals.

Whilst suggestive of a major shortcoming in the predictions of the
hierarchical model, the severity of this failure depends once
again on whether the luminosity-weighted ages probed can be
associated with the true formation epoch. An interpretation of our
data broadly consistent with the hierarchical framework would be
that most bulges are indeed produced from minor mergers of earlier
disk systems (and hence that the underlying stellar populations in
bulges are older than ellipticals) but that other physical
effects that are missing from the models also play an important
role in stimulating central star-formation after the bulk of the
bulge mass is already in place (eg. secular processes that
stimulate a modest amount of ongoing star formation in the
central regions).

Although our observations do not necessarily require a major
revision of the merger-driven astrophysics underlying the
hierarchical models, a significant puzzle is the greater scatter
in the colors of bulges compared to those in ellipticals (for
which similar merger-driven activities are presumably occurring).
The observed differences {\em might} be understood in terms of the
timescales involved (see discussion below), as well as by the
preferential morphological selection of ellipticals as relaxed
remnants (Menanteau et al 1999).

\subsection{Models Incorporating Secondary Star Formation}

The discussion in \S5.1 and \S5.2 suggests that the major issue
for understanding the ramifications of our optical observations
for {\em both} secular and hierarchical models is the
relationship between the luminosity-weighted ages of bulges and
their structural ages. The key to establishing this would seem to
be the measurement of the third ``age measure'' referred to
earlier: namely establishing the epoch of formation and mass
associated with the {\em initial} bulge-formation event.

In Figure~4 we show optical color-redshift trajectories for
passively-evolving stellar populations upon which we have
superposed 15\% of the stellar mass, either in the form of a
short-lived starburst or as an extended star-formation episode
(modeled as exponential star-formation with a 5 Gyr e-folding
timescale). The latter star-formation activity may be associated
with a quasi-continuous infall of gas. We illustrate models
corresponding to the onset of star-formation activity at
redshifts $z=0.2, 0.4, 0.6, 0.8,$ and $1.0$.

A notable feature of the optical data is the bimodal nature of the
deviations from the quiescent evolution track exhibited by the
early-type galaxy population. Starburst models do not remain blue
for much longer than the time associated with the active
star-formation phase. At high redshifts, our observations probe
sufficiently far into the ultraviolet that the observed $V$-band
flux is dominated by current star-formation even for a moderate
(10--15\% by mass) burst. Unsurprisingly a ``drip-feed'' scenario
results in a weaker but longer-lived blue phase suggesting that
our present data can be used to place crude constraints on the
duty-cycle of star-formation activity for both the bulges and
early-type galaxies in our sample. The locus of very blue
early-type systems combined with a paucity of intermediate-color
early-type systems suggests that star-formation in high-redshift
ellipticals is associated with rapid bursts whereas the broader
color distribution observed for the bulges may be suggestive of a
low-level, more extended star-formation.

As a quantitative example of the last point, consider two simple
models: a passively evolving 5 Gyr old stellar population with
roughly constant star-formation which then suffers a relatively
minor nuclear starburst, and a pure starburst model with no
underlying old stellar population. To be definitive, assume the
starburst has an exponential star-formation history with an
e-folding timescale of 100 Myr, and forms 10\% of the mass of the
underlying old population in the first model. Immediately after
the burst phase, the composite (old + young) stellar population
model at $z=1$ is only 0.12 mag redder in $V-I$ than the pure
burst model --- at high redshifts, the young stellar population
effectively masks the presence of any underlying old population in
optical bands. However, the corresponding figure for $I-H$ is 0.51
mag. Extending the present methodology to the infrared allows one
to determine, {\em in an individual case}, whether a blue bulge is
associated with initial bulge formation or incremental bulge
growth.

\subsection{Constraints from NICMOS Observations}

To make further progress we now turn to the resolved near-infrared
(IR) data made possible through NICMOS observations of the
HDF-North. As stated earlier, analysis of these data proceeded in
the same manner as the analysis of drizzled WF/PC2 optical data.
However, because the sampling of the NICMOS data is relatively
poor, care is needed to ensure precise co-registration of the
optical and IR data. Prior to our analysis all optical data was
first convolved with the $H_{160}$-band PSF and re-binned to
match the characteristics of the NIC3 dataset. The entire analysis
process described earlier was then repeated for the convolved and
rebinned optical data as well as for the NICMOS data (using, once
again, the $I_{814}$-band images to define all central
apertures). In order to test whether the coarser resolution of
the NIC3 data was a major source of error, we constructed a new
version of Figure~4 using the convolved and resampled $V_{606}$
and $I_{814}$ images, and found that the trends remained similar.

Even though only half our sample has been imaged (and with coarser
resolution) with HST in the infrared, the benefits NICMOS offers
are considerable. Firstly, disk contamination is expected to be
less important at longer wavelengths, so if we see the same
trends as in Figure 4, then this provides very strong
confirmation of differences in the evolutionary behaviour of
bulges and ellipticals as discussed above. Secondly, the addition
of the IR photometry can break some of the degeneracy in the
modelling based on optical colors discussed above since stars of
different main sequence timescales are involved. As emphasized
earlier, these advantages are offset by the poorer angular
resolution and sensitivity of the NICMOS data (as well as by
possible uncertainties in the modelling of stellar population
changes at near-IR wavelengths, as described by Charlot, Worthey,
\& Bressan 1996).

Figure 7 shows the \JH color-redshift relation for the HDF-N
targets plotted in Figure 4 with deep IR imaging. The model tracks
discussed earlier in the context of Figure 4 are also plotted for
the IR photometric bands. As mentioned, the IR models are mainly
used to indicate general trends and we have displaced the \JH BC96
model plotted in Figure~7 by +0.1 mag in order to ensure the
passive evolution track lies closer to the locus defined by the
redder E/S0 systems. We note that a \JH$=0.1$ mag color offset is
quite compatible with the systematic effects seen by Charlot,
Worthey, \& Bressan 1996, who investigated the agreement in the
predictions of independently-derived infrared models. Despite
these caveats, the qualitative benefit of IR models is clear;
timescales for synchronization of colours following a burst are
much shorter in the IR and, unlike the case with optical colours,
there is a very significant difference in the size of the color
excursions arising from secondary star formation depending upon
whether the bursts are intense and short-lived or of a gradual
``drip-feed'' nature.

With only half the HDF galaxies available, small number statistics
become a significant complication, and consequently we remain
cautious regarding the interpretation of the IR sample. However
the general impression from a comparison of Figures 4 and 7 is
that the optical/IR trends are very similar to intermediate
redshifts ($z\lesssim$0.6) but become somewhat different at high
redshift ($z\gtrsim0.6$).  At intermediate redshifts, in both
Figures~4 and 7 a dominant red E/S0 locus can be identified,
although a minority of the E/S0 population lies well blueward of
the overall relation. Clearly the photometric error is larger in
the IR data. Furthermore, for $z\lesssim$0.6, most of the spiral
bulges have optical and IR colors intermediate between these
extreme cases. However, at high redshifts, a significant fraction
of {\em infrared} bulges now lie alongside the red ellipticals,
and these are largely those in high central concentration spirals
(large open symbols). Indeed many $z\gtrsim0.6$ bulges appear to
be slightly {\em redder} than their elliptical counterparts.

Assuming (as justified earlier) that disk contamination does not
significantly complicate the interpretation of Figure 4 ($\S$3.2)
to redshifts $z\simeq$0.5 and that this is completely negligible
at all redshifts in Figure 7, then the combined data suggest that
bulge building activity at higher redshifts is likely to be
predominantly operating in a ``short-burst'' mode, rather than as
part of a more extended ``drip feed'' process. Comparison with the
model tracks suggests that the more extended recovery timescale
for optical colors following a low-mass, short burst could well
result in the majority of $z\gtrsim0.6$ bulges lying near the red
locus in the IR while at while being significantly offset in
optical color.

A more interesting interpretation, though tentative given the
small samples involved, would arise if the transition in the bulge
IR color distribution seen in Figure~7 at or around $z\simeq$0.6
is significant. This might imply that optically blue bulges are
more common at $z\lesssim0.6$ than at $z\gtrsim0.6$ which could be
understood if this redshift corresponded to one where a
fundamental change in the bulge-building history of spirals
occurred. This epoch has been claimed to be that at which barred
spiral systems become fairly common in deep imaging data (Abraham
et al. 1999). An increase in the fraction of blue bulges at this
redshift could provide further evidence that the formation of
Hubble's ``tuning fork'' occurs at or around this epoch, assuming
the excess of $z<0.6$ blue bulges corresponds to the onset of
secular bulge-building processes. This would strengthen the link
between bar formation and bulge-building (either through
bar-driven gas flows or bar dissolution). It is perhaps
interesting that a greater proportion of blue bulges at
$z\lesssim0.6$ occur in low central concentration galaxies than is
the case at $z\gtrsim0.6$, since the importance of secular
activity is probably a strong function of Hubble type, as
described earlier.

\subsection{Future Work}

The statistical relationship between bulge color and bulge-to-disk
ratio as a function of redshift may also yield information on the
epoch of initial bulge formation. If the structural parameters of
the bulge can be reliably disentangled from the underlying disk,
then high-redshift data that probes {\em in situ} formation from a
structural standpoint can be used to isolate the epoch of bulge
formation independently from the age of its stars. Such a study
would also allow us to extend the present work in a number of
interesting ways. In the present paper we have effectively
decoupled measures of galactic structure from measures of stellar
population, and hence we are unable to address quantitatively the
important question of whether the enhanced blue colors of spiral
bulges are associated with position on the Hubble sequence. Since
the best evidence from local data is that secular activity is most
closely associated with late-type spirals, studies linking the
relative colors of bulges and ellipticals with measures of growth
in bulge-to-disk ratio are required. Secular growth can also be
examined directly in high-redshift galaxy data by kinematical
observations that will soon become feasible with intermediate
resolution spectrographs on 10m-class telescopes. Such dynamical
data can be used to determine whether the bulk motions of
high-redshift bulges have a closer kinematical connection to disk
rotation or to thermal motions.

Coupling the present analysis to structural bulge-to-disk
decompositions will also allow investigations of the extent to
which the relatively blue bulges seen in our data may be the
manifestation of a colour-magnitude relation (CMR) in field
ellipticals, similar to that seen in well-studied cluster samples
(Sandage \& Visvanathan 1978). No CMR relation was found in high
redshift HDF field ellipticals by Kodama, Bower, \& Bell (1999),
although its presence may have been masked by uncertainties
arising from poorly-understood photometric redshifts.

Assuming that the slope of the field CMR at high redshifts
resembles that seen in local cluster data, then the CMR has a
completely negligible impact on our infrared bulge colors, and on
the optical colors at $z<0.5$. For example, the slope of the
$B-R$ CMR in Coma is $< 0.03$ mag$^{-1}$ (Milvang-Jensen \&
Jorgensen 2000), and presumably the $V-R$  slope is even smaller.
At higher redshifts, where the $I_{814}$ filter enters the
rest-frame $V$-band, the CMR may begin to make optical bulge
colors measurably bluer, although the effect probably represents a
relatively small contribution to the systematic error budget. For
example, in order for a Coma-like CMR to explain the whole of the
present trend for blue bulges would require the bulges in our
sample to be 6--8 mag fainter than typical ellipticals (assuming a
Coma-like CMR of $\Delta(B-V)$=0.05 mag$^{-1}$), which is clearly
incompatible with our visual pre-selection isolating relatively
early-type spirals with large bulge-to-disk ratios. Furthermore a
CMR is already incorporated into the hierarchical predictions
given in Figure 6. Nevertheless, it will be interesting to
compare the absolute magnitudes of the bulges in the present
sample with the corresponding absolute magnitudes for ellipticals
at similar redshifts, in order to establish the extent to which
cluster CMR data can be extended to spiral bulges in the field.

It is notable that the trends shown in Figures~4 and 7 are also
seen when 5\% apertures (instead of total apertures) are used to
define the colors of the E/S0 population. This is not surprising,
since Appendix~1 of Menanteau et al. (1999) indicates that the
mean metallicity of an elliptical interior to its effective
radius is reasonably close to the metallicity at the effective
radius. This calculation is also illuminating when assessing the
possible importance of metallicity gradients in biasing of the
colors of spiral bulges. If the metallicity gradients in bulges
are similar in strength to those in the luminous ellipticals
studied by Arimoto~et~al.~(1997), then the very slow drop-off in
the mean metallicity of light interior to an aperture larger than
the bulge effective radius implies that our 5\% apertures should
constitute a fair sample of the total bulge light.

Finally, we seek reassurance that the models outlined in Figures~4
and 7 are consistent with the colors and scatter observed locally
for bulges and ellipticals. Surprisingly, however, there is no
precise measure of the local homogeneity of bulges (Wyse et al
1997). Peletier et al (1999), in a particularly detailed
optical-infrared analysis of 20 nearby systems with HST, draw
attention to the redder colors of the inner 100-200 pc some
component of which they attribute to dust obscuration. (We note
that the presence of nuclear dust at this level, if confirmed,
would further strengthen the results of the present paper.) On
scales corresponding to bulge effective radii, where dust effects
are expected to be negligible, Peletier et al (1999) find mean
bulge colors comparable to those of Coma ellipticals. However,
imprecise photometric conversions were necessary to make this
comparison, rendering the precision of this statement no better
than $\pm$0.2 mag in $B-I$. The color dispersion in the bulges
themselves is typically also $\simeq$0.15-0.2 mag in $B-I$.
Clearly, improved local data are needed to establish the relative
colors of bulges and ellipticals in the local Universe.

\section{Conclusions}

A robust prediction of hierarchical scenarios that associates
morphological transformations with mergers and disk growth is that
bulges must be statistically redder and older than ellipticals. In
order to test this prediction, we have collated a magnitude
limited sample of spirals and ellipticals from both Hubble Deep
Fields and compared the aperture-based \VI color-redshift relation
for the central bulges of spirals selected in various ways with
those of field ellipticals.

In detail we find the following:

\begin{itemize}

\item At $z\lesssim0.6$, where we demonstrate that contamination from
underlying disk light is minimal, bulges are significantly bluer
than elliptical galaxies. Very few bulges follow
traditionally-accepted passive evolutionary tracks. At optical
wavelengths most bulges are 0.3-1.0 mag {\it bluer} than
ellipticals at comparable redshifts, with a significant color
dispersion at a given redshift, in marked contrast to the
expectations of recent semi-analytic predictions. Statistically
speaking, bulges remain optically bluer than ellipticals at all
redshifts $z\gtrsim 0.6$ to the limits of our data, and
contamination by bluer disk light and/or aperture mismatches is
unlikely to explain the bulk of this effect.

\item Although at first sight our data suggest that bulges
are recently formed, we show that secondary star formation
involving bursts of 15\% by mass superimposed upon a pre-existing
old stellar population can readily reproduce the color dispersion
observed. We discuss two general scenarios: short-lived bursts and
extended events possibly associated with continuous infall of gas.

\item Near-infrared NICMOS observations of a sub-sample of our dataset
agree with the optical trends at $z\lesssim 0.6$, but at higher
redshifts most infrared bulges appear to be as red as most
ellipticals. Model comparison with the optical trends suggests
that the duty cycle for the optical blueing of high-redshift
bulges is quite short. Most importantly, however, our data firmly
rule out the traditional picture whereby bulges form at high
redshift and evolve passively as miniature ellipticals. Although
somewhat speculative at this stage, the divergence observed in
the IR properties of bulges and ellipticals at $z\sim 0.6$ may
suggest that other physical processes become important in forming
galactic bulges, for example secular evolution in stellar disks
linked to the formation of galactic bars.

\item Much work needs to be done at both high and low
redshifts in order to establish the star-formation history of
bulges in various galaxy types and their environments. It is clear
that such studies are central to testing hierarchical models for
the morphological evolution of galaxies. In this paper we have
shown significant differences between the distribution of colors
in  bulges and ellipticals which either seriously challenge our
theoretical views on bulge formation or, more likely, reveal the
necessity for including additional physics into these models,
corresponding to periods of extended star formation in the inner
regions of spiral galaxies.

\end{itemize}

\section{Acknowledgements}

We thank Carlton Baugh, Shaun Cole and Carlos Frenk for their
enthusiastic support of this project and Carlton Baugh for
generously giving us semi-analytic predictions in the context of
our observations. We thank Stephen Gwyn for kindly making his
photometric catalog available to us, and for matching the object
identifications in his catalog to those given in the STScI HDF
catalog. Reynier Peletier is thanked for making the results from
his NICMOS bulge progam known to us in advance of publication. We
acknowledge useful discussions with Jarle Brinchmann, Ray
Carlberg, Roger Davies, Gerry Gilmore, Piero Madau, Ron Marzke,
Felipe Menanteau, Mike Merrifield, Rachel Somerville and Chuck
Steidel.


\newpage
\begin{figure}[ht]
  \centering
\caption{{\bf SEE JPEG IMAGE: F1.JPG}. $I_{814}$-band images of
all non-peculiar, non-blended spirals in our sample (Plate~1). The
small yellow central aperture shown on each galaxy is used to
determine the bulge color as described in the text. Redshifts and
elliptical-aperture-based total $I_{814}$-band magnitudes are
shown on each panel. Photometric redshifts are shown in
parentheses.}
  \label{montageFig}
\end{figure}

\newpage
\begin{figure}[ht]
  \centering
   \epsfig{figure=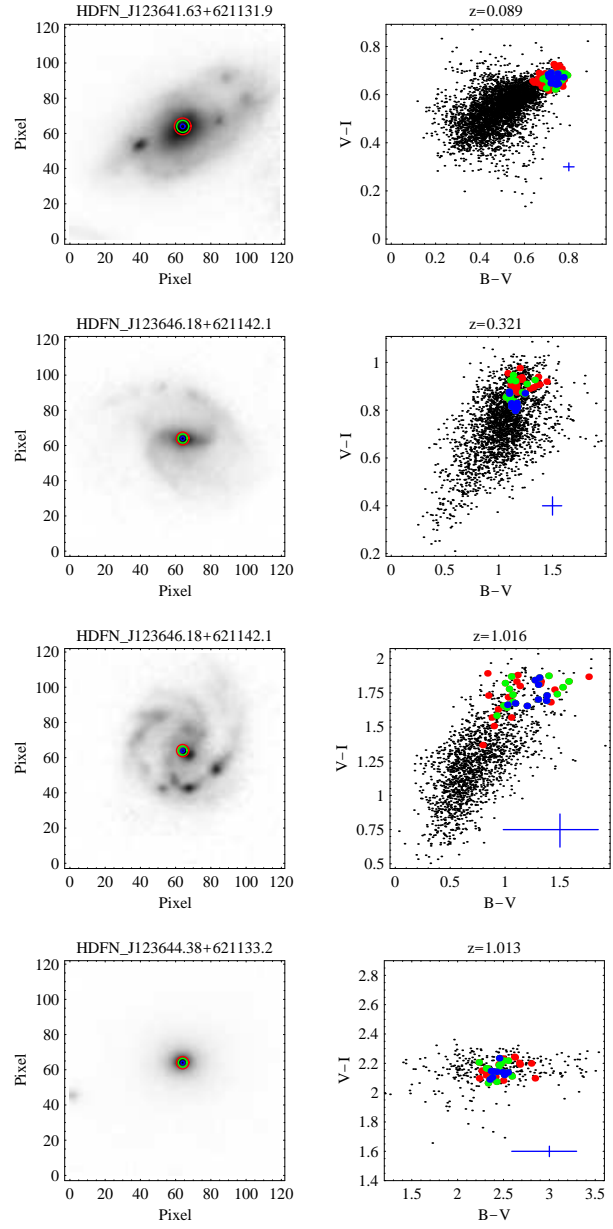,height=7in}
\caption{[Left] $I_{814}$-band images of three bright spirals and
one elliptical from our sample. Circles correspond to fractional
radial apertures of 2\% (blue), 5\% (green) and 7\% (red) of the
semi-major axis length of the galaxy. [Right] Corresponding
pixel-by-pixel color-color distributions. Pixels internal to the
circular apertures on the left-hand panels are shown with the same
color coding. The error bars shown correspond to those for
typical pixels in the 2\% apertures.}
  \label{montageFig}
\end{figure}

\newpage
\begin{figure}[ht]
  \centering
  \epsfig{figure=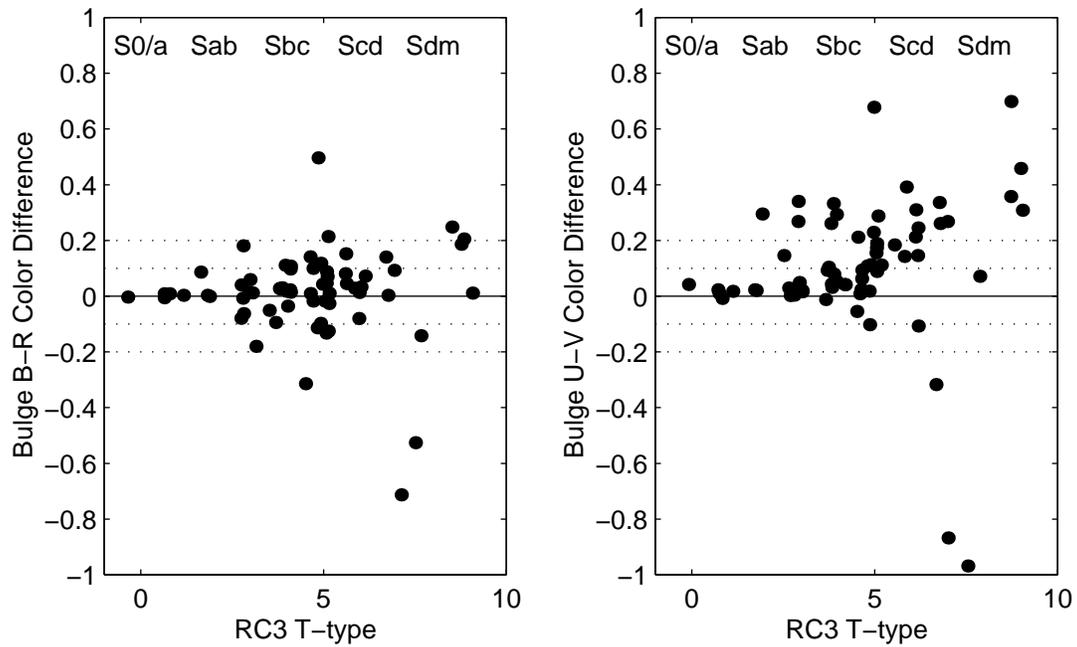}
\caption{The difference between bulge colors determined using the
simple aperture-based  procedure described in the text and bulge
colors determined using two-dimensional fits to galaxy surface
brightness distributions, for the nearby sample of spirals of
de~Jong (1996). Color differences are shown as a function of
Hubble sequence $T$-type from the {\em Third Reference Catalog}.
For clarity, $T$-types are plotted with small random offsets. The
difference between the estimated $B-R$ and $U-V$ bulge colors are
shown in the left and right panels, respectively. As described in
the text, these approximate to the rest-frame colors corresponding
to \VI at redshifts of 0.4 and 0.7, respectively.}
\label{deJongFig}
\end{figure}

\newpage
\begin{figure}[ht]
  \centering
\epsfig{figure=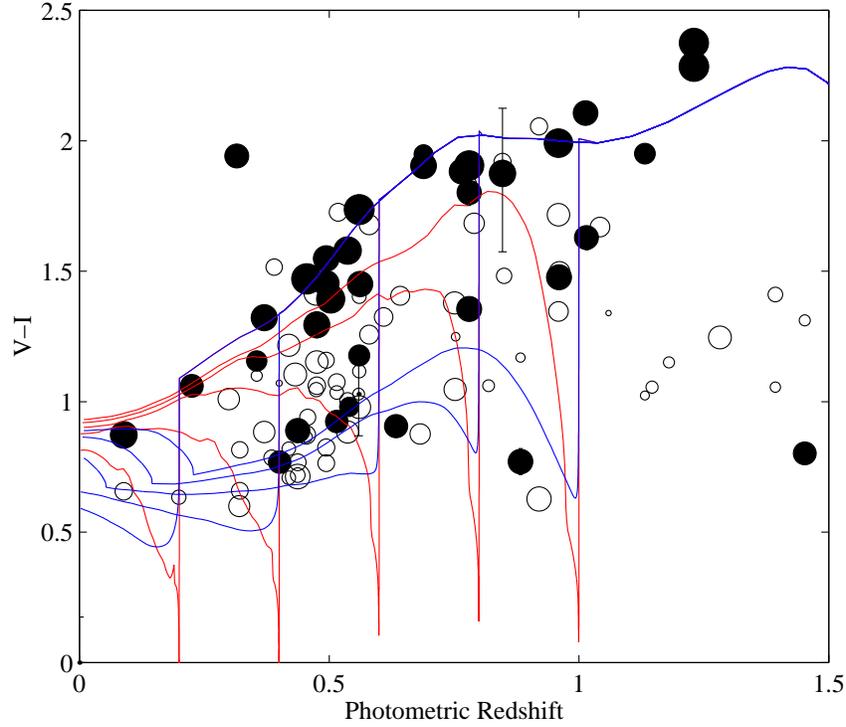,width=4.5in}
\caption{\VI~colors for ellipticals (solid symbols, large
aperture) and spiral bulges (open symbols, 5\% aperture - see
text) as a function of photometric redshift. The spiral sample was
defined according to the entries in Table 1 with RGA=Sab|Sbc|Disc,
blend=no|minor, bulge=yes, pec=no. Symbol sizes are proportional
to galactic central concentration, as defined in Abraham et al.
(1996).  Representative error bars are shown on several data
points. Note the striking offset in the colors of spiral bulges
relative to the red elliptical locus at any given redshift. This
offset is much larger than the expected contribution from blue
disk light shown in Figure~\ref{deJongFig}. The solid blue curve
at the top represents the observed \VI color expected for a
passively-evolving system that formed in a single burst at $z$=3.
This is the baseline model upon which secondary bursts are added.
The other curves define color-redshift trajectories for systems
which suffer secondary bursts of activity at redshifts of
0.2,0.4,0.6,0.8 and 1. Red curves refer to 0.1 Gyr bursts
involving 15\% of the stellar mass whereas the blue curves refer
to extended bursts with 5 Gyr e-folding timescales (see \S5.3 in
the text for details). }
  \label{EBulgeFig}
\end{figure}

\newpage
\begin{figure}[ht]
    \centering
    \epsfig{figure=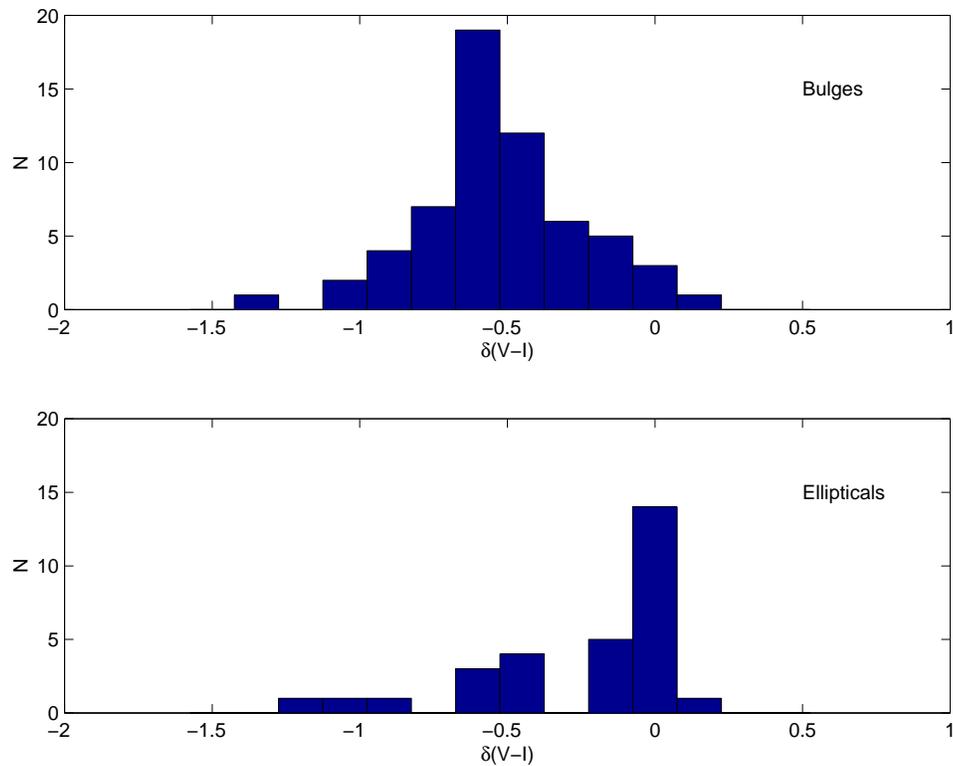,width=5in}
    \caption{The distribution of \VI colors for the spiral bulges
and ellipticals plotted in Figure 4 with respect to the passive
evolutionary prediction. Although these color differences span
similar ranges in the observer's frame, the distributions are
quite different with the elliptical distribution showing bimodal
characteristics.}
    \label{offsetsig}
\end{figure}

\newpage
\begin{figure}[ht]
  \centering
  \epsfig{figure=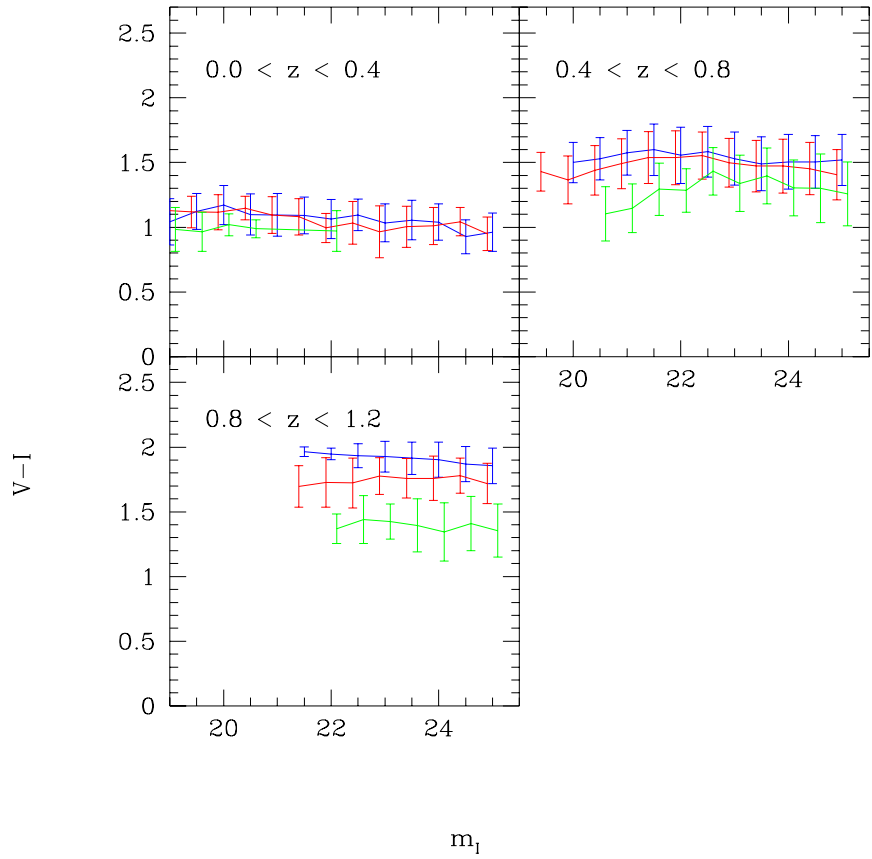,width=5in}
  \caption{The predicted colors (courtesy of Carlton Baugh) of bulges,
ellipticals, and S0 galaxies in three broad redshift bins:
$0<z<0.4$ (top left), $0.4<z<0.8$ (top right), and $0.8<z<1.2$
(lower left) based upon the semi-analytical prescription of Baugh
et al. (1999). \VI~color versus total galaxy $I_{814}$ magnitude
is shown for spiral bulges (blue), ellipticals (red) and S0
galaxies (green). Mean colors are joined by solid lines, and
error bars correspond to the color variance.}
  \label{predictionFig}
\end{figure}

\newpage
\begin{figure}[ht]
    \centering
    \epsfig{figure=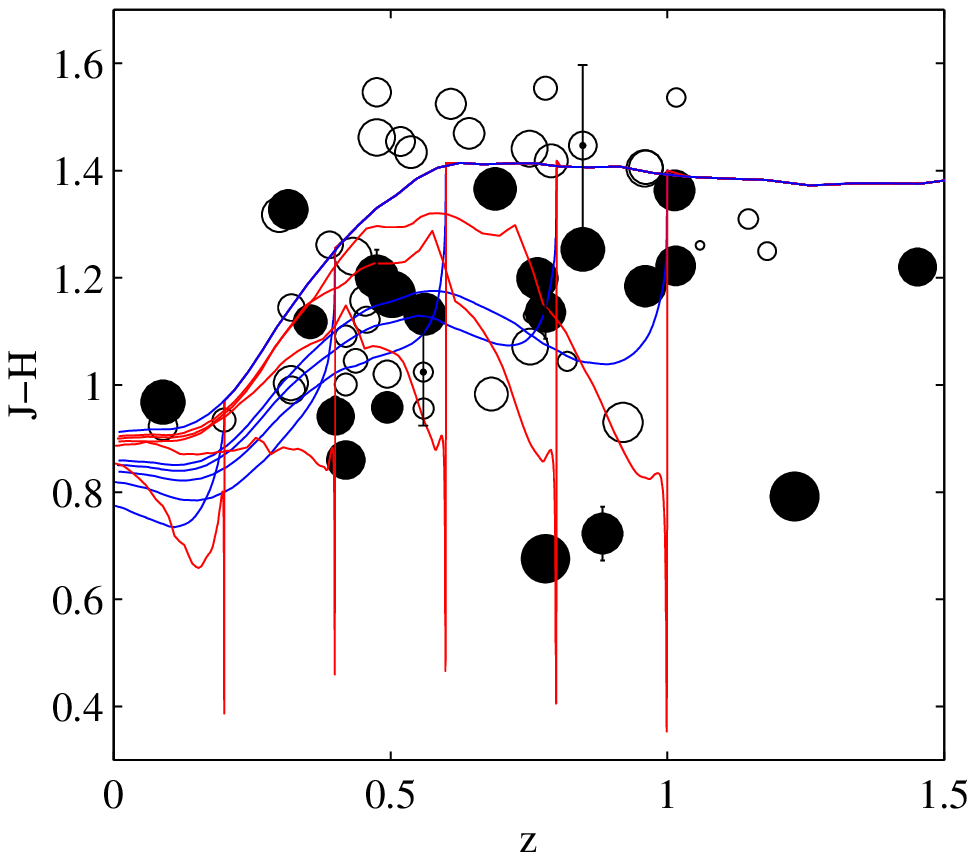,width=5in}
    \caption{Infrared J-H color redshift distribution for the HDF
sample of bulges and ellipticals adopted in Figure 4 with the same
symbols and model predictions (see text for details).}
    \label{ircolorsfig}
\end{figure}


\begin{thebibliography}{}

\bibitem[]{} Abraham, R. G., Tanvir, N. R., Santiago, B. X., Ellis, R. S.,
Glazebrook, K. \& van  den Bergh, S. 1996, MNRAS, 279, L47

\bibitem[Abraham et al. (1999a)]{rga99a} Abraham, R.G., Merrifield,
M., Ellis, R.S., Tanvir, N.R. \& Brinchmann, J. 1999a {\it Mon.
Not. R. astr. Soc.} in press (astro-ph/9811476).

\bibitem[Abraham et al. (1999b)]{rga99b} Abraham, R.G., Ellis,
R.S., Fabian, A.C., Tanvir, N.R. \& Glazebrook, K. 1999b {\it
Mon. Not. R. astr. Soc.}, {\bf 303}, 641.

\bibitem[Andredakis \& Sanders (1994)]{as94} Andredakis, Y. C. \& Sanders, R. H.
1994, {\it Mon. Not. R. astr. Soc.}, {\bf 267}, 283.

\bibitem[]{} Arimoto, N., Matsushita, K., Ishimaru, Y., Ohashi, T., \&
Renzini, A. 1997, ApJ, 477, 128

\bibitem[Baugh et al. (1998)]{baugh98} Baugh, C.M., Cole, S., Frenk,
C.S., Lacey, C.G. 1998 {\it Astrophys. J.}, {\bf 498}, 504.

\bibitem[Bower et al. (1998)]{bower98} Bower, R.G., Kodama, T. \&
Terlevich, A. 1998 {\it Mon.  Not. R. astr. Soc.}, {\bf 299},
1193.

\bibitem[Brinchmann et al. (1998)]{jarle98} Brinchmann, J., Abraham,
R.G., Schade, D.J., Tresse, L., Ellis, R.S., Lilly, S.J.,
LeFevre, O., Glazebrook, K., Hammer, F. \& Colless, M. 1998 {\it
Astrophys. J.}, {\bf 499}, 112.

\bibitem[Bunker (1999)]{bunker99} Bunker, A. in {\em The OCIW Workshop
on Photometric Redshifts}, ed. Weymann, R. et al, in press
(astro-ph/9907196).

\bibitem[Charlot \& Bruzual 1999]{charlot99} Charlot, S. \& Bruzual, G. 1999,
in preparation.

\bibitem[Combes (1999)]{combes99} Combes, F. 1999 preprint (astro-ph/9904031)

\bibitem[de~Jong(1996)]{dj96} de Jong, R. 1996 {\it Astr. Astrophys.}, {\bf 313}, 377.

\bibitem[Dickinson(2000)]{med00} Dickinson, M.E. 2000 in {\it
Building Galaxies: From the Primordial Universe to the Present},
XIXth Moriond Astrophysics Meeting, eds. Hammer, F. et al,
(Paris:Ed. Frontieres), p257.

\bibitem[Driver, Windhorst, \& Griffiths (1995)]{driver95} Driver, S.,
Windhorst, R.A. \& Griffiths, R.E. 1995 {\it Astrophys. J.}, {\bf
453}, 48.

\bibitem[Dwek (1995)]{dwek95} Dwek, E. et al 1995 {\it Astrophys. J.} {\bf 445}, 716.

\bibitem[Eggen, Lynden-Bell, \& Sandage (1962)]{els} Eggen, O.J.,
Lynden-Bell, D. \& Sandage, A. 1962 {\it Astrophys. J.}, {\bf
136}, 748.

\bibitem[Frankston \& Schild (1976)]{fs76} Frankston, M., \&
Schild, R. (1976), {\it Astron. J.}, {\bf 81}, 500

\bibitem[Kauffmann, White, \& Guiderdoni (1993)]{guin93}
Kauffmann, G., White, S.D.M. \& Guiderdoni, B. 1993 {\it Mon.
Not. R. astr. Soc.}, {\bf 264}, 201.

\bibitem[Kauffmann \& Charlot (1998)]{guin98} Kauffmann, G. \& Charlot, S. 1998 {\it Mon. Not. R. astr. Soc.}, {\bf 297}, L23.

\bibitem[Kodama et al (1999)]{kodama99} Kodama, T., Bower, R.G. \& Bell,
E.F. 1999 {\it Mon. Not. R. astr. Soc.}, (in press).

\bibitem[Kormendy (1993)]{kor93} Kormendy, J. 1993, in {\em Galactic Bulges},
eds. Dejonghe, H. and Jabing, H. J, Kluwer Academic Publishshers,
p.209-228

\bibitem[Kuijken \& Merrifield (1995)]{km95} Kuijken, K. \& Merrifield,
M. 1995 {\it Mon. Not. R.  astr. Soc.}, {\bf 443}, L13.

\bibitem[Lilly et al. (1998)]{lilly98} Lilly, S.J., Schade, D.J.,
Ellis, R.S., LeFevre, O., Brinchmann, J., Abraham, R.G., Tresse,
L., Hammer, F., Crampton, D., Colless, M. \& Glazebrook, K. 1998
{\it Astrophys. J.}, {\bf 500}, 75.

\bibitem[Marleau \& Simard (1998)]{ms98} Marleau, F. \& Simard, L. 1998
{\it Astrophys. J.}, {\bf 507}, 585.

\bibitem[Marzke et al. (1994)]{ron98} Marzke, R.O. et al. 1994 {\it
Astron. J.}, {\bf 108}, 437.

\bibitem[Menanteau et al. (1999a)]{felipe99a} Menanteau, F., Ellis,
R.S., Abraham, R.G., Barger, A. \& Cowie, L. 1999a {\it Mon. Not.
R. astr. Soc.}, in press (astro-ph/9811465)

\bibitem[Menanteau, Abraham, \& Ellis (1999b)]{felipe99b} Menanteau,
F., Abraham, R.G. \& Ellis, R.S. (2000), in preparation.

\bibitem[Milvang-Jensen & Jorgensen (2000)]{mjj00}
Milvang-Jensen, B. \& Jorgensen, I. (2000); astro-ph/0004049

\bibitem[Norman \& Sellwood (1996)]{norm96} Norman, C., Sellwood, J. \& Hasan, H. 1996 {\it Astrophys. J.}, {\bf 462}, 114.

\bibitem[Peletier et al. (1999)]{pel99}Peletier, R.F., Balcells, M., Davies, R.L., Andredakis, Y., Vazdekis, A., Burkers, A. \& Prada, F. 1999 {\it
Mon. Not. R. astr. Soc.}, submitted.

\bibitem[Ratnatunga, Griffiths, \& Ostrander]{rat99} Ratnatunga, K.U.,
Griffiths, R.E. \& Ostrander, E.J.  1999 {\it Astron. J.}, in
press (astro-ph/9904179)

\bibitem[Rich (1997)]{rich97} Rich, M. 1997 in {\it The Central Regions of the
Galaxy and Galaxies}, IAU Symposium 184, Kluwer.

\bibitem[Schade et al. (1995)]{schade95} Schade, D.J., Lilly, S.J.,
Crampton, D. et al 1995, {\it Astrophys. J.}, {\bf 451}, L1.

\bibitem[Schade et al. (1996)]{schade96} Schade, D.J., Crampton, D.,
Hammer, F., LeFevre, O.  \& Lilly, S.J. 1996 {\it Mon. Not. R.
astr. Soc.}, {\bf 278}, 95.

\bibitem[Schade et al. (1999)]{schade99} Schade, D.J., Lilly, S.J.,
Crampton, D., Ellis, R.S., LeFevre,O., Hammer, F., Brinchmann, J.,
Abraham, R.G., Colless, M.  \& Tresse, L. 1999 {\it Astrophys.
J.}, in press (astro-ph/9906171).

\bibitem[Sellwood (1999)]{sellwood99} Sellwood, J. 1999 in {\it Galaxy
Dynamics}, eds. Merritt, M. et al, in press (astro-ph/9904084)

\bibitem[van~den~Bergh (1998)]{vdb98} van den Bergh, S. 1998, ``Galaxy
Morphology and Classification'', (Cambridge: Cambridge University
Press).

\bibitem[Williams~et~al. (1996)]{wil96} Williams, R. et al 1996 {\it Astron. J.}, {\bf 112},
1335.

\bibitem[Williams~et~al. (1999)]{wil99} Williams, R. et al 1999 in preparation.

\bibitem[ Wyse, Gilmore, \& Franx]{wgf97} Wyse, R.F.G., Gilmore, G. \&
Franx, M. 1997 {\it Ann.  Rev. Astron. Astrophys.}, {\bf 35}, 637.

\end{thebibliography}
\end{document}


\begin{deluxetable}{lcccccccccccccccc}
\tabletypesize{\small}
\tablecaption{HDF North and South Sample}
\tablehead{
  \colhead{ID\tablenotemark{a}} &
  \colhead{$I_{814}$\tablenotemark{b}} &
  \colhead{$H_{160}$\tablenotemark{c}} &
  \colhead{$z$\tablenotemark{d}} &
  \colhead{$V-I$\tablenotemark{e}} &
  \colhead{$V-I$\tablenotemark{f}} &
  \colhead{$J-H$\tablenotemark{g}} &
  \colhead{$J-H$\tablenotemark{h}} &
  \colhead{$\epsilon$\tablenotemark{i}} &
  \colhead{$C$\tablenotemark{j}} &
  \colhead{$A$\tablenotemark{k}} &
  \colhead{Auto\tablenotemark{l}} &
  \colhead{RSE\tablenotemark{m}} &
  \colhead{RGA\tablenotemark{n}} &
  \colhead{bulge?\tablenotemark{o}} &
  \colhead{blend?\tablenotemark{p}} &
  \colhead{pec?\tablenotemark{q}}
}
\startdata
HDFN\_J123649.44+621346.9   &             18.19         &             16.10         &             0.089         &             0.86          &             0.87          &             1.22          &             0.97          &             0.33          &             0.68          &             0.06          &             Early         &              0            &             E             &             yes           &             no            &             no            \\
HDFS\_J223254.90-603144.1   &             18.75         &             \nodata       &             -1.000        &             1.86          &             1.84          &             \nodata       &             \nodata       &             0.08          &             0.90          &             0.20          &             Star          &             -2            &             Star          &             no            &             no            &             no            \\
HDFS\_J223305.04-603400.8   &             19.18         &             \nodata       &             -1.000        &             1.71          &             1.64          &             \nodata       &             \nodata       &             0.09          &             0.89          &             0.30          &             Star          &             -2            &             Star          &             no            &             no            &             no            \\
HDFN\_J123656.35+621241.2   &             19.20         &             17.03         &             -1.000        &             0.58          &             0.68          &             0.72          &             0.83          &             0.19          &             0.90          &             0.14          &             Star          &             -2            &             Star          &             no            &             no            &             no            \\
\enddata
\tablenotetext{a}{IAU designation from STScI SExtractor-based catalogs}
\tablenotetext{b}{SExtractor ${\rm mag}_{best}$ from STScI catalogs}
\tablenotetext{c}{$H$-band total magnitude}
\tablenotetext{d}{Redshift (photometric redshift in parentheses). Spectroscopically confirmed stars are denoted by $z=-1.000$}
\tablenotetext{e}{$V-I$ Color for central 5\% of galaxy (see text for details)}
\tablenotetext{f}{$V-I$ Color of total galaxy based on elliptical apertures defined from $I$-band image moments}
\tablenotetext{g}{$J-H$ Color for central 5\% of galaxy (see text for details)}
\tablenotetext{h}{$J-H$ Color of total galaxy.}
\tablenotetext{i}{1.0 - axial ratio of galaxy}
\tablenotetext{j}{Central concentration}
\tablenotetext{k}{Asymmetry}
\tablenotetext{l}{Automated classification}
\tablenotetext{m}{Ellis numerical classification}
\tablenotetext{n}{Abraham visual classification}
\tablenotetext{o}{Flag indicating the presence of a well-defined bulge (based on visual inspection)}
\tablenotetext{p}{Flag indicating degree of object blending (based on visual inspection)}
\tablenotetext{q}{Flag indicating clear galaxy peculiarity aside from blending/superposition}
\end{deluxetable}

\begin{deluxetable}{lccccccc}
\tabletypesize{\small}
\tablecolumns{8} \tablecaption{Non-peculiar, unblended
subsets from $I_{814}<24$ mag sample} \tablehead{
  \colhead{}    &  \multicolumn{3}{c}{Photometric Z} &   \colhead{}   &
  \multicolumn{3}{c}{Spectroscopic Z} \\
  \cline{2-4} \cline{6-8} \\
  \colhead{Description of subset} &
  \colhead{$N$} &
  \colhead{$\langle\delta(V_{606}-I_{814})\rangle$\tablenotemark{\dag}} &
  \colhead{$\sigma_{\delta(V_{606}-I_{814})}$} &
  \colhead{} &
  \colhead{$N$} &
  \colhead{$\langle\delta(V_{606}-I_{814})\rangle$\tablenotemark{\dag}} &
  \colhead{$\sigma_{\delta(V_{606}-I_{814})}$}
} \startdata \sidehead{Early type\tablenotemark{\star}}
Unblended, early-type galaxies  &    36 & -0.083 &  0.448 & &    11 & -0.136 &  0.435\\
Unblended, early-type galaxies (RGA visual classification\tablenotemark{\ddag} ) &    36 & -0.083 &  0.488 & &    12 & -0.171 &  0.464\\
Unblended, early-type galaxies (RSE visual classification) &    47 & -0.220 &  0.490 & &    19 & -0.212 &  0.467\\
Unblended elliptical galaxies (RGA classed as E,E/S0) &    26 & -0.075 &  0.504 & &     9 & -0.101 &  0.318\\
\sidehead{Late type\tablenotemark{\star}}
Unblended late-type galaxies with obvious bulges  &    48 & -0.526 &  0.330 & &    24 & -0.487 &  0.284\\
Unblended late-type galaxies with obvious bulges (RGA visual classification\tablenotemark{\ddag} ) &    65 & -0.544 &  0.309 & &    29 & -0.474 &  0.284\\
Unblended late-type galaxies with obvious bulges (RSE visual classification) &    47 & -0.538 &  0.288 & &    19 & -0.433 &  0.204\\
Unblended late-type galaxies with obvious bulges and area $>$ 2000 pix  &    36 & -0.480 &  0.226 & &    19 & -0.431 &  0.211\\
\enddata
\tablenotetext{\dag}{Median central $\delta(V_{606}-I_{814})$ as
defined in the text.} \tablenotetext{\star}{Based on automated
classifications unless specified otherwise.}
\tablenotetext{\ddag}{Plotted in Fig.4}
\end{deluxetable}